# OPTIQU: COORDINATED MULTI-LEVEL VOLTAGE AND REACTIVE POWER CONTROL FOR ENHANCED VOLTAGE QUALITY AND SECURE GRID OPERATION

Irene Hammermeister[1*], Eric Tönges[2], Nils Bornhorst[2], Johannes Heid[1,2], Gabriela Fritzler[1], Timo Rehwald[1], Andrea Schoen[1], Ronald Halbauer[3], Jan Meschede[3], Michael Kramer[4], Julia Holl[4], Sina Straußberger[4], Andrey Luzhbin[5], Maximilian Niedhammer[5], Mischa Geiger[5], Aaron Eicker[6], Maurice Raetsch[6], Alfio Spina[6], Johannes Dieplinger[7], Christian Mayer[7], Josef Bayer[8], Thorsten Reske[9], Dominik Hilbrich[1]

*[1]Fraunhofer Institute for Energy Economics and Energy System Technology IEE, Kassel, Germany*
*[2]University of Kassel, Kassel, Germany*
*[3]Mitteldeutsche Netzgesellschaft Strom mbH, Kabelsketal, Germany*
*[4]Thüga AG, München, Germany*
*[5]Thüga Energienetze GmbH, Schifferstadt, Germany*
*[6]TU Dortmund University, Dortmund, Germany*
*[7]Maschinenfabrik Reinhausen GmbH, Regensburg, Germany*
*[8]EnSolVision GmbH, Parsberg, Germany*
*[9]Dipl.-Ing. H. Horstmann GmbH, Heiligenhaus, Germany*
**irene.hammermeister@iee.fraunhofer.de*

## Abstract

Modern low-voltage (LV) distribution grids face rising shares of photovoltaic generation and high-power loads such as heat pumps and electric vehicle charging stations. Due to high simultaneity, voltage constraints often become binding before thermal limits, triggering costly conventional grid reinforcement measures. Existing voltage and reactive power control in LV grids – e.g., fixed cos(φ) or Q(V) control of distributed generators, on-load tap-changing distribution transformers, and line voltage regulators – is typically applied locally and independently, leaving reactive power flexibility potential unused. This paper presents OptiQU, a coordinated voltage and reactive power control concept for medium-voltage (MV) and LV distribution grids, combining centralised optimisation with decentralised local control and fallback strategies. The approach coordinates operational targets and setpoints across MV and LV (e.g., DER reactive power and substation equipment) to mitigate voltage violations and curtailment and to increase hosting capacity, while enabling robust operation under limited communication. The concepts are being evaluated using representative MV/LV models in simulation and lab environments and will be validated in field tests with two German DSOs. Based on existing research, the coordinated approach is expected to increase the exploitable flexibility for upstream voltage and reactive power control. The planned evaluation will quantify this potential and investigate trade-offs between performance, communication effort, and resilience.

## 1 Introduction

Voltage management in distribution grids is increasingly shaped by interactions across multiple voltage levels. The rapid growth of inverter-based distributed energy resources (DER), particularly photovoltaic systems, and the electrification of heat and mobility significantly increase voltage variability and simultaneity effects in medium-voltage (MV) and low-voltage (LV) grids. Several studies indicate that voltage constraints in LV grids are frequently reached before thermal limits, thereby restricting hosting capacity for additional DER and electrified loads [1] [2]. While MV and LV grids are directly affected by DER and electrified demand, operators must simultaneously maintain local voltage limits and satisfy reactive power requirements at higher voltage interfaces [3] [4]. In current operation, voltage and reactive power control in MV and LV grids is predominantly local and uncoordinated, leaving available reactive power flexibility of inverter-based generation, regulated transformers and secondary-substation equipment insufficiently exploited [4].
This paper addresses the question of how decentralised voltage and reactive power control capabilities in MV and LV grids can be coordinated to support upstream reactive power requirements while ensuring secure and resilient DSO operation.

The main contributions of this paper are:

- a coordinated MV/LV voltage and reactive power control concept, referred to as OptiQU, enabling the joint activation of reactive power resources and transformer tap positions to maintain voltage limits and support upstream reactive power requirements;
- a conceptual framework, within OptiQU, combining optimisation and decentralised fallback control, designed for practical DSO operation and resilient response to potential communication outages;
- an evaluation framework covering simulation, laboratory, and field-oriented validation with German DSOs.

The remainder of this paper is organised as follows. Section 2 introduces the coordinated control concept, optimisation approach and system architecture. Section 3 presents the evaluation setup and expected results. Section 4 concludes and outlines implications for DSO operation and future work.

## 2 Methodology

To address the coordination challenge outlined in the Introduction, the OptiQU methodology for coordinated MV/LV voltage and reactive power control is structured in three parts. It first introduces the typical MV/LV grid structure – including controllable assets – and the relevant coordination use case in Section 2.1. They form the foundation for the optimisation and control strategies presented in Section 2.2. It then describes the applied optimisation approaches, and finally outlines the system architecture in Section 2.3, which enables centralised coordination, decentralised control, and resilient fallback operation.

### *2.1 MV/LV System Overview and Coordination Use Case*

*2.1.1 System Scope and Controllable Assets:*
The OptiQU methodology considers coordinated voltage and reactive power control in typical MV and LV distribution grids. A representative MV feeder is supplied by an HV/MV substation and comprises multiple MV/LV substations feeding radial LV networks with inverter-based distributed energy resources (DER), electrified loads, and conventional demand.
Within this MV/LV structure, coordinated voltage and reactive power control requires assets that can influence voltage levels and reactive power exchange across voltage levels. The following controllable assets are therefore considered:

- DER, whose reactive power setpoints can be directly controlled by the DSO at the MV and LV levels,
- inverter-based distributed generators operating with Q(V), Q(P) or conventional cos(φ) control at the LV level,
- OLTC-equipped transformers at HV/MV and MV/LV substations,
- secondary-substation controllers enabling setpoint execution and local fallback strategies.

In addition, intelligent measurement systems (e.g., smart meters and digital secondary substations) accompanied by distribution system state estimation provide the multi-level grid observability required for coordinated operation.
OLTCs integrated into MV/LV voltage-regulating distribution transformers (VRDTs) provide a practical means of maintaining voltage quality in increasingly dynamic LV grids. By adjusting the transformer turn ratio in discrete steps, the voltage setpoint – typically defined on the low-voltage side – can be stabilised under varying load and generation conditions.
By influencing the voltage at the MV/LV interface, OLTCs indirectly affect the reactive power behaviour of LV-connected DER operating with voltage-dependent characteristics (e.g., Q(V)). While OLTC-equipped transformers partially decouple the medium- and low-voltage levels under varying operating conditions, they provide a controllable interface between both levels. Within this framework, OLTCs serve as key local actuators, responding to local setpoints or signals from higher-level algorithms. In this way, they add an additional degree of freedom for coordinated reactive power control across MV and LV grids.
The increasing deployment of digital secondary substations and intelligent measurement infrastructure enhances both controllability and observability in MV and LV grids. Together with inverter-based DER and OLTC-equipped transformers, these assets form the technical basis for the coordinated use case described in Section 2.1.2.

*2.1.2 Multi-Level Coordination Use Case:*
To illustrate the OptiQU approach, a multi-level coordination use case is considered (Fig. 1), comprising an upstream controller at the EHV/HV interface that defines a reactive power target at the HV/MV interface and the OptiQU controller at HV/MV level, which coordinates MV and LV reactive power flexibility to meet this target while respecting local voltage constraints and asset limits. The coordination process consists of four main steps:
**Step 1** – The OptiQU controller at the HV/MV interface determines the aggregated feasible reactive power range ($Q$min, $Q$max) of controllable MV and LV assets, including inverter-based generators and OLTC-equipped transformers.
**Step 2** – The upstream controller at the EHV/HV interface aggregates this flexibility with HV-connected resources to determine the overall reactive power range.
**Step 3** – Based on the aggregated flexibility, the upstream controller computes reactive power targets and setpoints to meet TSO requirements and HV voltage limits. The target for the HV/MV interface is communicated to the OptiQU controller.
**Step 4** – The OptiQU controller allocates the HV/MV reactive power target within the flexibility ranges determined in Step 1 and applies direct control by transmitting setpoints to inverter-based generators and OLTC-equipped MV/LV transformers. OLTC tap changes enable indirect control of LV-connected assets through voltage-dependent characteristics (e.g., Q(V)), thereby shaping the overall reactive power response.

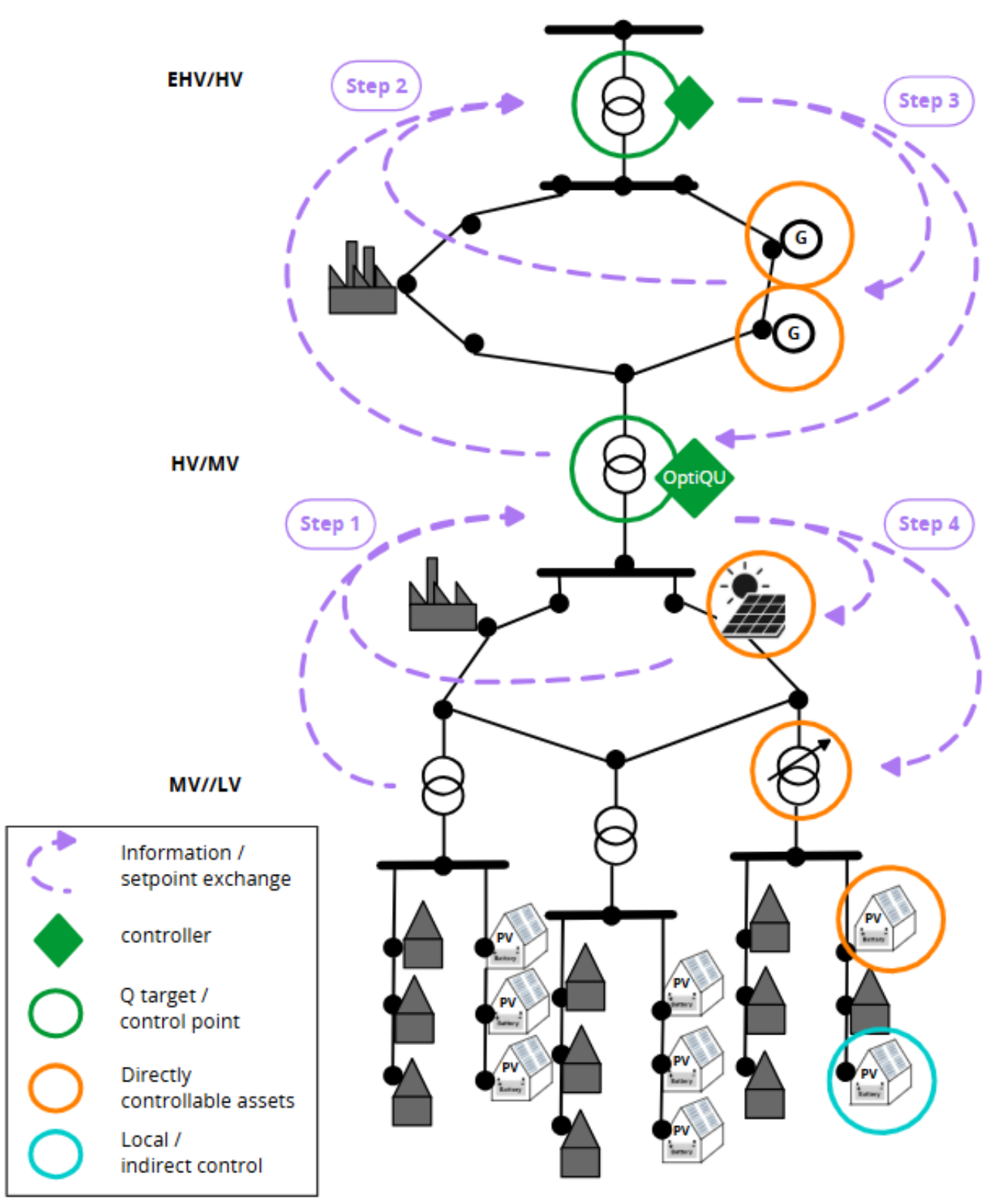

Figure 1: Coordinated multi-level voltage and reactive power control with sequential flexibility assessment

*2.2 Optimisation Approach*

The optimisation processes based on the use cases described in Section 2.1 require suitable optimisation approaches that provide both result quality and scalability for application during grid operation. In general, the flexibility determination and setpoint calculation in each step are subject to physical power flow constraints (nonlinear) and operational bounds. Thus, the optimisation problem can be expressed as an AC optimal power flow (OPF) problem, as described in [5], with several modifications described below.

Within the OptiQU controller, this optimisation problem is used with different objective functions depending on the task: reactive power flexibility aggregation and setpoint determination.

In the flexibility aggregation (Figure 1, Step 1), the resulting reactive power flow over the considered transformer is minimised and maximised in two distinct optimisations to capture the entire range of reactive power flexibility. In the setpoint calculation step, the objective function aims to minimise the deviations from the reactive power targets requested by the upstream controller (Figure 1, Step 3), while staying within operational limits.

During the optimisation, only reactive power setpoints of controllable loads and generators and transformer tap positions are variables to be controlled, while active power setpoints remain constant. Since transformer setpoints are discrete, the overall optimisation problem is a non-convex mixed-integer programming problem, which is computationally challenging. For continuous problems, interior-point solvers such as ipopt [6] provide locally optimal solutions, and approaches as described in [7] allow for the iterative solution of mixed-integer AC OPF problems.

While these approaches based on mathematical optimisation yield (locally) optimal, high-quality solutions, scalability for real-world instances is questionable. Therefore, runtime-efficient alternative approaches such as [8] are adapted to the specifications described. The AC OPF is used as a benchmark to evaluate result quality of the sensitivity-based method, other heuristic approaches and linearisations of the AC OPF problem. To increase efficiency, an additional aggregation level at the MV/LV interface is considered, limiting the optimisation to single voltage levels each.

Figure 2 depicts the overall structure of optimisation approaches applied to the use case from Section 2.1.

In case of communication failure, decentralised approaches are required to prevent violations of local voltage limits. This can be realised by either pre-defined fall-back profiles for controllable assets, or by decentralised devices placed at OLTCs. The latter allows for the application of the outlined optimisation logic considering local measurements even in cases where no central optimisation device is available and ensures operation within operational limits.

Problem formulation
Solution Approaches
Evaluation
Nonconvex AC OPF
Objectives:
Q-flexibility, Q-setpoints
Mathematical
Accuracy (Benchmark)
Sensitivity-based
Scalability
Performance Metrics
(runtime,
result quality)

Figure 2: Coordinated multi-level voltage and reactive power optimisation approach

*2.3 System Architecture*

The effectiveness of the optimisation strategies described in Section 2.2 depends critically on a system architecture capable of deploying, monitoring, and executing setpoints across the MV/LV grid. In OptiQU, this is achieved through a modular system framework that integrates central optimisation, decentralised local control, and multi-level measurement data. The system architecture (Section 2.3.3) includes a central coordination platform (Section 2.3.1) for optimisation and setpoint distribution, as well as comprehensive measurement and data sources (Section 2.3.2) spanning MV and LV grids. This architecture ensures resilient operation, secure communication, and scalable integration of existing DSO infrastructure.

*2.3.1 Coordination Platform – beeDIP:*

beeDIP, developed by Fraunhofer IEE, is a platform for deploying and testing pilot applications in real-world grid operation without interfering with existing control systems [9]. It enables the integration of external software modules – such as a reactive power optimiser corresponding to the approach described in Section 2.2 – and harmonises heterogeneous data sources through a unified data model. This supports the modular, scalable and resilient architecture of OptiQU and facilitates the secure evaluation of new control approaches in MV/LV grids. beeDIP is based on a microservices architecture, enhancing robustness and scalability. A graphical user interface is provided for system configuration, control and monitoring.

A two-tier architecture is considered in OptiQU: The central optimiser (Section 2.2), implemented on the beeDIP platform, calculates setpoints based on aggregated flexibility potentials and distributes them to decentralised control instances deployed on edge devices – local computing units at the secondary substations that execute optimisation and control functions close to the assets. Each device receives a sub-problem to meet the assigned setpoint while respecting local constraints and solving it autonomously. The local control itself is a containerised microservice, resulting in a hardware-independent implementation. It is evaluated whether the beeDIP framework can be deployed on the edge devices to ensure homogeneity of the system architecture. Alternatively, containerised microservices can be deployed directly to the device. In both configurations, the deployed microservices act as the interface between the central optimisation and subordinate grids. Critically, the decentralised controllers maintain fallback autonomy: even when communication with the central optimiser is lost, local control ensures safe and resilient operation. Further, the edge devices can take on subtasks of the central optimisation and therefore provide additional resilience to the system.

*2.3.2 Measurement and Data Sources:*

OptiQU relies on comprehensive measurement data across MV and LV grids to enable coordinated voltage and reactive power control. Existing DSO infrastructure provides real-time data from digital secondary-substations, intelligent measurement devices, and smart meters. Measurements include voltages, currents, reactive and active power flows, and asset status information.

The beeDIP-based OptiQU server platform (Section 2.3.1) accesses these data via the DSOs' existing cloud solutions, ensuring a consolidated view of the grid state while maintaining operational security. It also ensures continuity and reliability of measurements and metadata, supporting consistent evaluation and control strategies even if some sources are temporarily unavailable.

*2.3.3 System Overview:*

This section illustrates how the OptiQU system integrates its server platform, edge devices, and measurement infrastructure to realise coordinated MV/LV control. In the OptiQU concept, the term 'server platform' refers to beeDIP (Section 2.3.1), which hosts the reactive power optimisation described in Section 2.2 as a microservice. Figure 3 shows the overall OptiQU system architecture.

At field level, measurements and controllable assets are located at the HV/MV substation, multiple MV/LV intelligent secondary substation nodes (iSSN) and the connected LV feeders. Operating data from intelligent secondary substations and the smart metering infrastructure (via SMGW(A)) are transmitted to cloud data platforms, where monitoring functions (data ingestion, time-series storage and visualisation) are separated from the operational optimisation service. The cloud platform exchanges measurements and data with the beeDIP-based OptiQU server platform via a REST API or directly via an InfluxDB API. The central operational optimisation computes coordinated MV setpoints to satisfy MV constraints and HV/MV requirements such as a reactive power target $Q_{target}$ at the HV/MV substation. It then derives reactive power setpoints for directly controllable MV-connected DER and reactive power setpoints for each MV/LV substation. At each MV/LV substation, an edge device runs a cascaded local optimisation and control that determines the MV/LV transformer tap position and translates the substation setpoint into LV-level control actions, including indirect voltage-based control and, where available, direct reactive power setpoints for LV-connected DER, while respecting local LV constraints. Setpoints are distributed through a lightweight messaging layer (e.g., MQTT) to the edge devices deployed at MV/LV substations enabling resilient fallback behaviour if connectivity to the cloud layer is degraded or lost.

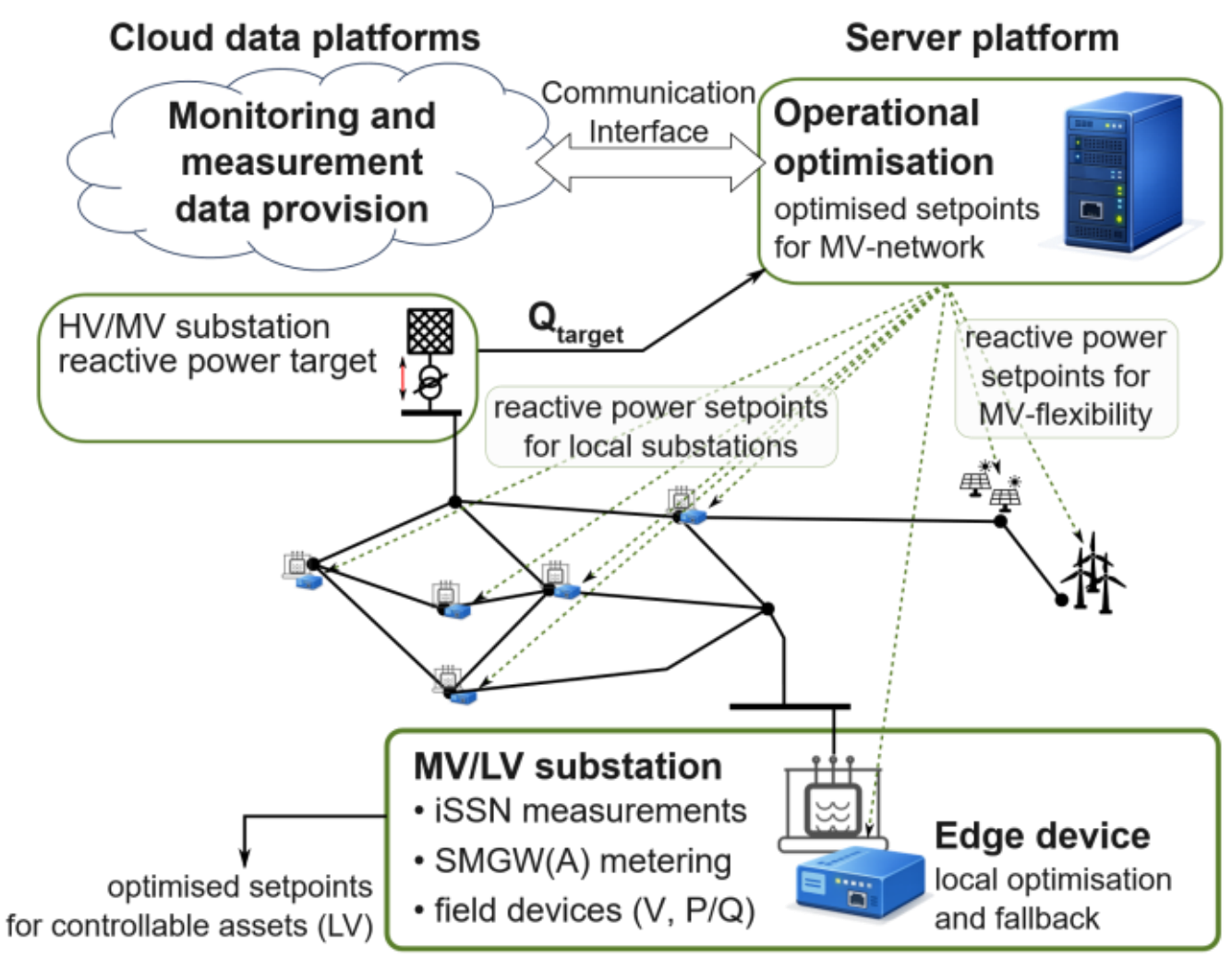

Figure 3: Overall OptiQU system architecture

## 3 Evaluation Setup and Expected Results

### *3.1 Evaluation Setup*

This section describes the evaluation framework of the OptiQU concept, including test grids, simulation-to-field validation, and expected outcomes.

#### *3.1.1 Test Grids and Scenarios:*

The assessment is based on anonymised real-world MV/LV grids provided by participating German DSOs, with three representative test grids. The primary MV feeder is directly connected to a 115/20 kV, 40 MVA HV/MV transformer and comprises 15 secondary substations with connected PV systems, wind turbines, storage units, a charging park, commercial loads, and about 160 households. Several substations are digitally equipped, and an OLTC-equipped MV/LV transformer enables the analysis of coordinated tap and reactive power control. The additional grids focus on LV integration and are suitable for testing in-line tap changers and OLTCs as a flexibility potential for the OptiQU controller. Historical measurements, supporting model parameterisation and algorithm development, are mainly available for MV grids, while LV grids are currently less observable. Synthetic worst-case scenarios (e.g., high demand with low renewable generation) complement historical data to stress-test operating conditions. During the field test phase, real-time measurements from the selected MV feeder will validate the coordinated control concept under actual operating conditions.

#### *3.1.2* Validation chain:

The anonymised datasets serve as the basis for detailed distribution grid modelling. The assessment is conducted using a power-hardware-in-the-loop (PHIL) simulation, in which selected inverter-based assets relevant for voltage and reactive power control are physically reproduced in the laboratory environment. Available assets include PV inverters, charging infrastructure, and battery storage systems. The remaining grid components are implemented in a real-time simulator, which is coupled to the laboratory power grid via a power amplifier [10], [11]. Required communication channels between the central and decentralised controllers, as well as the relevant assets, are emulated to enable the reception of measurement data and the transmission of setpoints. This laboratory setup is validated to ensure consistency between simulation and field conditions. All concepts and algorithms are validated using this laboratory setup before they are deployed in the field.

### *3.2 Expected Outcome and Discussion*

The OptiQU concept is expected to demonstrate how coordinated MV/LV voltage and reactive power control enhances the utilisation of existing distribution grid assets and increase effective hosting capacity without requiring immediate grid reinforcement. Key outcomes include quantifying the available reactive power flexibility in MV and LV grids, assessing the relative impact of controllable assets at different voltage levels, and analysing their interaction with transformer tap operations. Additionally, OptiQU evaluates trade-offs between control effectiveness, communication effort, and operational robustness, supporting resilient voltage management under realistic DSO constraints.

# 4 Conclusion and Outlook

The OptiQU concept presents a coordinated MV/LV voltage and reactive power control approach addressing the limitations of purely local, uncoordinated control schemes. By combining reactive power provision and transformer tap operations, the approach aims to improve voltage stability and optimise the use of existing assets.
Key recommendations for DSOs include prioritising the utilisation of existing inverter-based resources and OLTC-equipped transformers over conventional grid reinforcement, designing control architectures that ensure resilient operation with secure fallback strategies, and aligning coordinated voltage management with long-term stability requirements.
Further work will evaluate coordinated MV/LV voltage and reactive power control under different operating scenarios, identifying where the approach delivers the highest technical and economic benefit, supporting the targeted selection of feeders and secondary substations for the deployment of OLTC-equipped MV/LV transformers and other voltage control measures.
Both centralised and decentralised implementations will be tested, including hardware-in-the-loop setups, and a structured evaluation framework will assess effectiveness, risks, opportunities, and costs to guide practical deployment.

# 5 Acknowledgements

This paper presents results from the research project *OptiQU*, funded by the Federal Ministry for Economic Affairs and Climate Action (BMWK) under grant number 03EI4099 within the 8th Energy Research Program “Research Missions for the Energy Transition”.
AI-based tools were used for linguistic refinement and for creating selected graphical icons. All technical content and results are the responsibility of the authors.

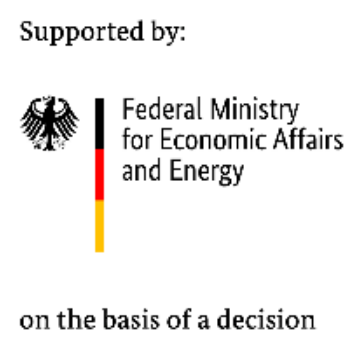